\documentclass[reqno]{amsart}
\usepackage{graphicx,amsmath, amsaddr}
\usepackage{float}
\usepackage[utf8]{inputenc}
\usepackage{color}
\usepackage{cite}
\usepackage{fancyhdr}
\usepackage[margin=2cm]{geometry}
\usepackage{appendix}
\usepackage{multirow}
{ 
	\sloppy

\title{Transient non-Fourier behavior of large surface bodies}

\author{R. Kovács}
\address{
	$^1$Department of Energy Engineering, Faculty of Mechanical Engineering, Budapest University of Technology and Economics, Műegyetem rkp. 3., H-1111 Budapest, Hungary \\
	$^2$Department of Theoretical Physics, Wigner Research Centre for Physics, Institute for Particle and Nuclear Physics, Budapest, Hungary \\
	$^3$Montavid Thermodynamic Research Group, Budapest, Hungary
}
\date{\today}

\begin{document}

	\maketitle

\begin{abstract}
The variety and complexity of heterogeneous materials in the engineering practice are continuously increasing, open-cell metal foams filled with phase change materials are typical examples. These are also having an impact on the recent developments in the energy industry. Earlier room temperature heat pulse experiments on macroscale foam samples showed non-Fourier over-diffusive behavior on a particular time scale. Since there is a need to investigate such complex structures on larger spatial scales and extend the one-dimensional analysis on two-, and three-dimensional settings, here we develop a two-dimensional analytical solution for the Guyer-Krumhansl and Jeffreys-type heat equations in cylindrical coordinates to investigate the transient thermal behavior of large bodies. We provide the steady-state and transient temperature and heat flux distributions for a space-dependent heat source. The solutions presented here will be helpful for the thermal characterization of complex materials and for the validation of numerical methods.

\end{abstract}
\vspace{0.5cm}

\maketitle

\section{Introduction}
Numerous experimental and theoretical studies emerged on room-temperature heat conduction beyond Fourier in recent years. On the one hand, the nanoscale effects result in the deviation from Fourier's law, usually with the appearance of ballistic heat conduction \cite{Maj93, Chen00, Chen02}. Furthermore, the size dependence of thermal conductivity enjoys great interest as it significantly influences the effectiveness of any nanoscale device \cite{WangGuo10a, AlvJou08}. On the other hand, room temperature non-Fourier heat conduction is not restricted to the nanoscale exclusively, and it is observable in macroscopic bodies under various conditions \cite{Botetal16, ChenZan09, NazmEtal21}. While the parallel diffusive and ballistic propagation modes are present on a nanoscale, the macroscopic deviation is due to the interaction of multiple parallel diffusive (and additional heat transfer) channels. 
Typical examples are rocks \cite{FehEtal21} and foams \cite{FehKov21, LunEtal22}. Although each component behaves according to Fourier's law, the heterogeneous material structure overall (effectively) leads to a more complex, non-Fourier temperature history. That was the motivation for two-temperature models \cite{Sobolev94, Sobolev16, Sobolev97}.

\begin{figure}[H]
	\centering
	\includegraphics[width=15cm,height=7cm]{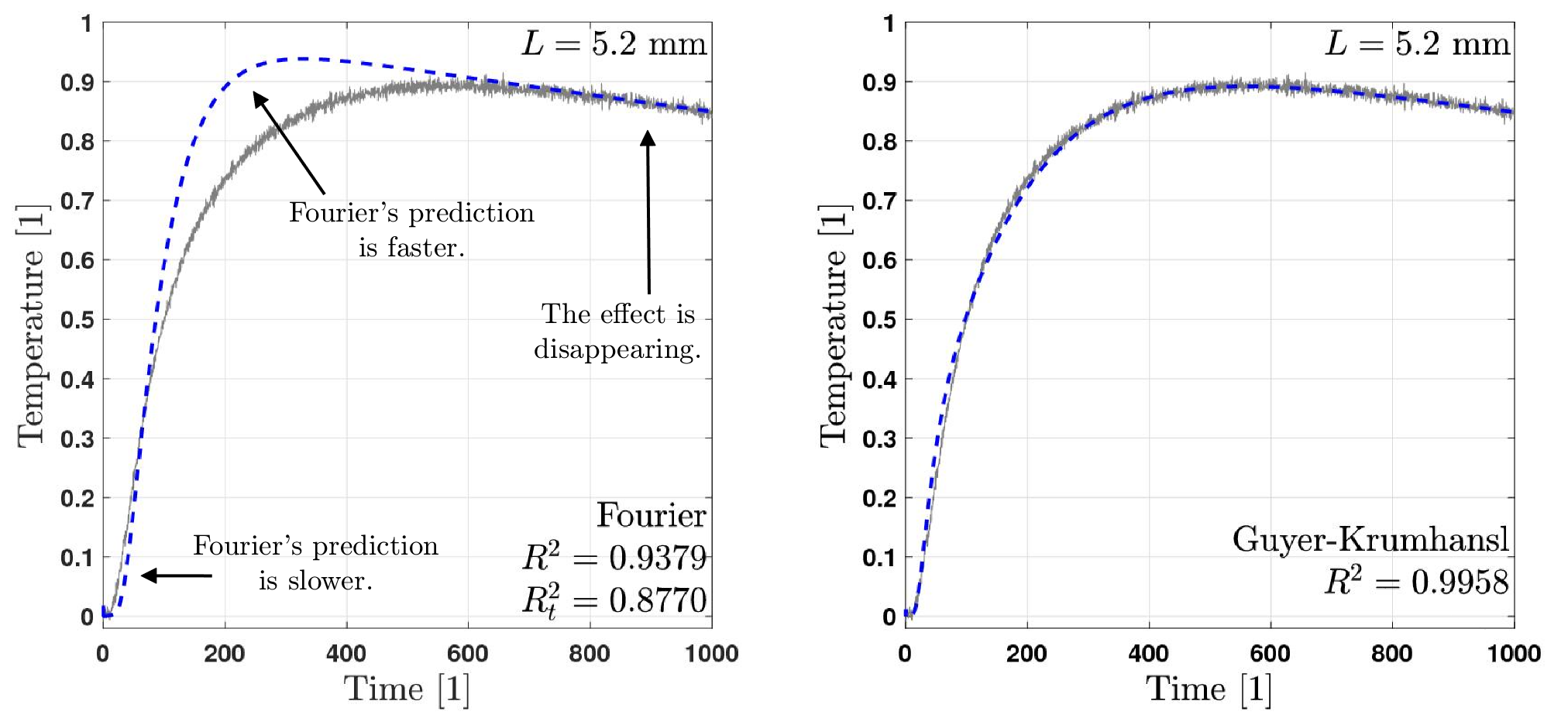}
	\caption{Typical appearance of over-diffusion \cite{FehKov21}.}
	\label{fig1}
\end{figure}

The presence of multiple time scales is the most visible by showing experimental data obtained from a heat pulse experiment for an aluminum foam (Figure \ref{fig1}) possessing multiple heat transfer channels. The response for a short but finite single pulse ($0.01$ s), together with the best achievable Fourier fit, shows that at least two heat conduction time scales are present simultaneously. This is called over-diffusion \cite{Tzou95, Botetal16} and so far best modeled with the continuum Guyer-Krumhansl (GK) heat equation \cite{FehKov21}. Here, with the word "continuum" is an adjective, referring to the continuum thermodynamic background of the GK equation \cite{VanFul12}, it is free from the usual kinetic theory and phonon hydrodynamic assumptions, therefore it is valid on much larger temperature and spatial scales, independently of the Knudsen number. It is also worth noting that in Fig.~\ref{fig1}, the time is re-scaled with respect to the pulse length, i.e., the dimensionless time $\hat t = t/0.01 $. The transients become slow enough after about $7$ s, meaning that the heat transfer process needs much more time to cancel out the effect of multiple time scales. In other words, if thermal transients continuously occur in some particular application, Fourier's law might not apply. Otherwise, such an experiment reveals the limitations on time scales, and the Guyer-Krumhansl equation provides a refinement of the thermal parameters in order to cover the faster processes as well. However, this property indeed scales with the size (different heat capacities, heat transfer surfaces, time scales) and surface (boundary conditions), and thus it is necessary to extend the experimental and theoretical capabilities in this direction.

It is worth noting that recent heat exchanger applications exploit the advantageous properties of a metal foam structure: having large heat transfer surfaces, the matrix material is an excellent heat conductor, therefore such solutions can essentially ease the realization of an effective thermal storage method. One outstanding example is when an open-cell foam structure is filled with phase change material \cite{ChenEtal21, NematEtal22, ZhangEtal22, WangEtal23}. The phase change materials usually have low thermal conductivity, significantly restricting their melting or solidification properties. However, a surrounding foam structure can notably enhance the thermal behavior, thus both the heating and cooling processes can be much more efficient. This further motivates the present study as there are currently no reliable thermal models which enable the resource-friendly modeling of such complex structures. A non-Fourier model, however, can be exceptional when the role of the parallel heat transfer channels is understood correctly in such an approach. The present study aims to take a step forward in this direction, deepening our understanding and extending our modeling possibilities about the Guyer-Krumhansl heat equation. In the following, let us briefly summarize the heat conduction models we consider here.

The well-known Fourier law is
\begin{align}
	\mathbf q = - \lambda \nabla T, \quad \lambda \in \mathbb{R^+} \label{eq1}
\end{align}
together with the balance of internal energy ($e=c_v T$),
\begin{align}
	\rho c_v \partial_t T + \nabla \cdot \mathbf q = q_v(\mathbf x, t), \label{eq2}
\end{align}
includes only one time scale, described by the thermal diffusivity $\alpha = \lambda /(\rho c_v)$, where $\lambda$, $\rho$ and $c_v$ are the thermal conductivity, mass density, and isochoric specific heat, respectively. Furthermore, $\mathbf{q}$ and $T$ stand for the heat flux and temperature fields, and $q_v$ is an internal heat generation, which could be time and space-dependent. We restrict ourselves to isotropic rigid materials. 

For a non-Fourier heat conduction model, the constitutive equation, Eq.~\eqref{eq1}, is exchanged with a more general expression, usually consisting of additional time and space derivatives. In the present paper, we consider the following two constitutive equations among the various models. First, we study the Guyer-Krumhansl  equation,
\begin{align}
	\tau \partial_t 	\mathbf q+	\mathbf q = - \lambda \nabla T + \eta_1 \Delta 	\mathbf q + \eta_2 \nabla \nabla \cdot 	\mathbf q, \quad \lambda, \tau, \eta_1, \eta_2 \in \mathbb{R^+}, \label{eq3}
\end{align}
in which $\tau$ is the relaxation time; $\eta_1$ and $\eta_2$ are independent intrinsic length scales, not associated with a propagation mechanism in a continuum model \cite{VanFul12}. We note that in the conventional treatment of the GK equation, $\eta_1 = l^2$ with $l$ being the mean free path of phonons, and $\eta_2/\eta_1=2$ for the particular approximations performed by Guyer and Krumhansl \cite{GK66}. We emphasize that for a macroscopic room temperature problem, the phonon approach is not valid anymore, while the continuum model, although possessing the same structure, is free from any prior specific assumptions on the propagation mechanism, hence extending the model's domain of validity. 

Second, we will continue our analysis with the Jeffreys equation (JE), i.e., 
\begin{align} 
	\tau_q \partial_t 	\mathbf q+	\mathbf q = - \lambda \nabla T - \lambda \tau_T \partial_t \nabla T, \quad \lambda, \tau_q, \tau_T \in \mathbb{R^+} \label{eq4}
\end{align}
where, instead of introducing further spatial derivatives, two time lags appear similarly to the popular dual-phase-lage (DPL) concept \cite{Tzou95}. However, while no thermodynamic background is behind the DPL model, Eq.~\eqref{eq4} can be derived on a thermodynamic basis \cite{SzucsEtal21}. Although we remain in the linear regime, it is worth noting that the coefficients are not completely independent of each other (also for Eq.~\eqref{eq3}) in a sense that the $T$-dependence of $\lambda$ would influence all the other parameters, too \cite{KovRog20}. For the Jeffreys equation, $\tau_q$ and $\tau_T$ can be adjusted almost independently, and the only exception is that when $\tau_q = 0$, $\tau_T=0$ follows immediately (but not vice versa). The GK and JE models consist of two time scales in different ways, and they share the same $T$-representation in a one-dimensional setting. However, their physical basis is quite different, and the GK equation fits much better into the systematic structure of non-Fourier models. 

In this paper, we choose to study large surface bodies since the typical heterogeneous materials also show strong size-dependent behavior \cite{FehEtal21}. In other words, extending the existing flash experiment for much larger bodies will be necessary as the usual thickness limit is about $3-6$ mm for standardized equipment. This can be much smaller than the representative sample size for a heterogeneous material, especially for foams with large (3-5 mm) open-cell structures. Furthermore, we aim to investigate both models using an analytical solution for a two-dimensional setting, which will emphasize the structural differences between these equations. We note that even the book of Carslaw and Jaeger \cite{CarJae59b} has limitations towards the problem setting we discuss in the following. 

For simplicity and clarity, we start and present our method on the example of the Fourier heat equation. This will provide an insight into the problem setting and the solution method. We will continue with the GK and JE equations, demonstrating how the solution method is applied to more complicated models. Such analytical solutions, especially for the GK equation, cannot be found in the literature. Moreover, as there is also missing a reliable two or three-dimensional numerical method, we offer a good starting point for future studies in this direction. Finally, we will compare the temperature histories to the Fourier equation and investigate whether we find the transient behavior similar (or even the same) compared to the one-dimensional room temperature experiments on small samples.

\section{Problem statement}
\begin{figure}[H]
	\centering
	\includegraphics[width=14cm,height=8cm]{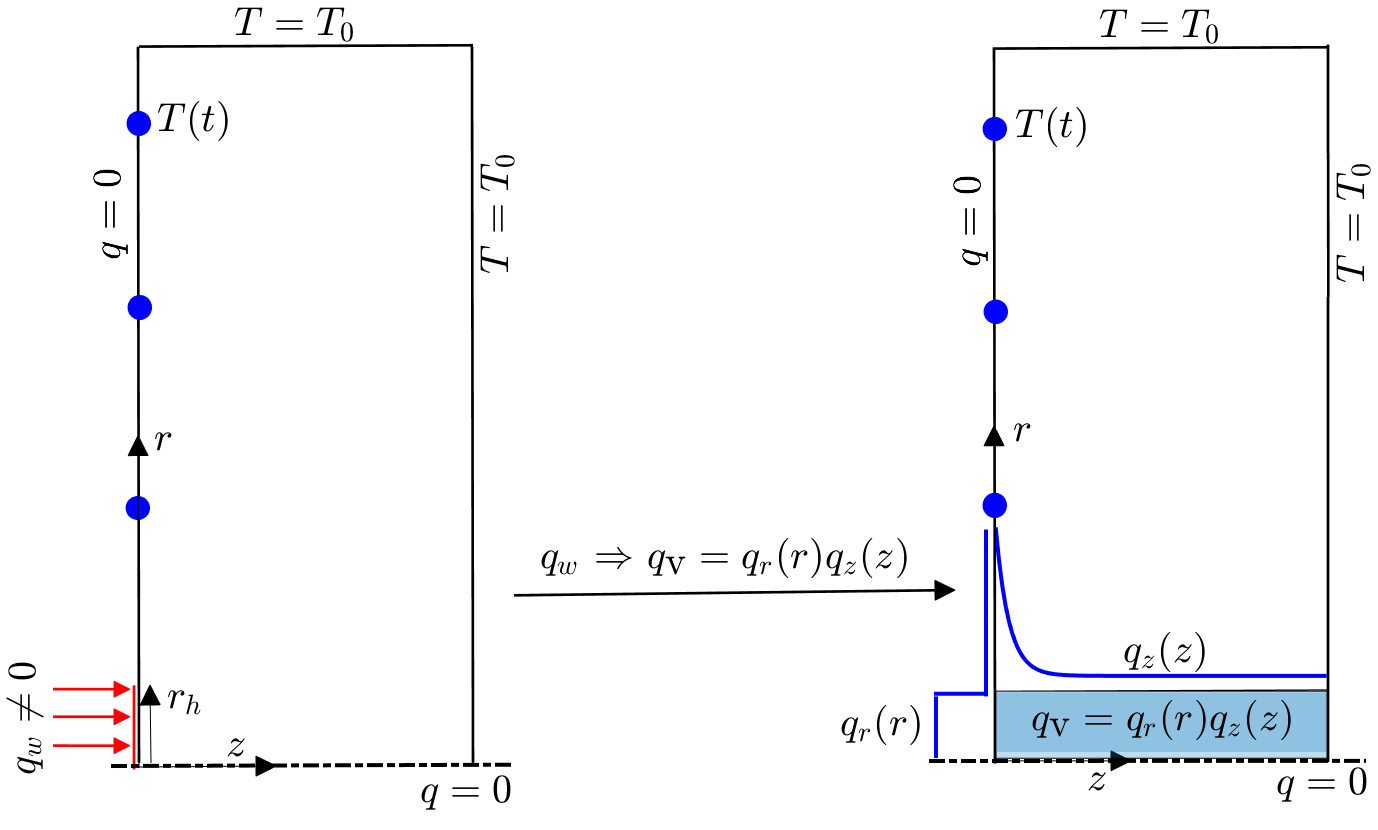}
	\caption{Schematic problem setting, presenting the boundary conditions and the characteristics of internal heat generation.}
	\label{fig2}
\end{figure}

\begin{figure}[H]
	\centering
	\includegraphics[width=11cm,height=6cm]{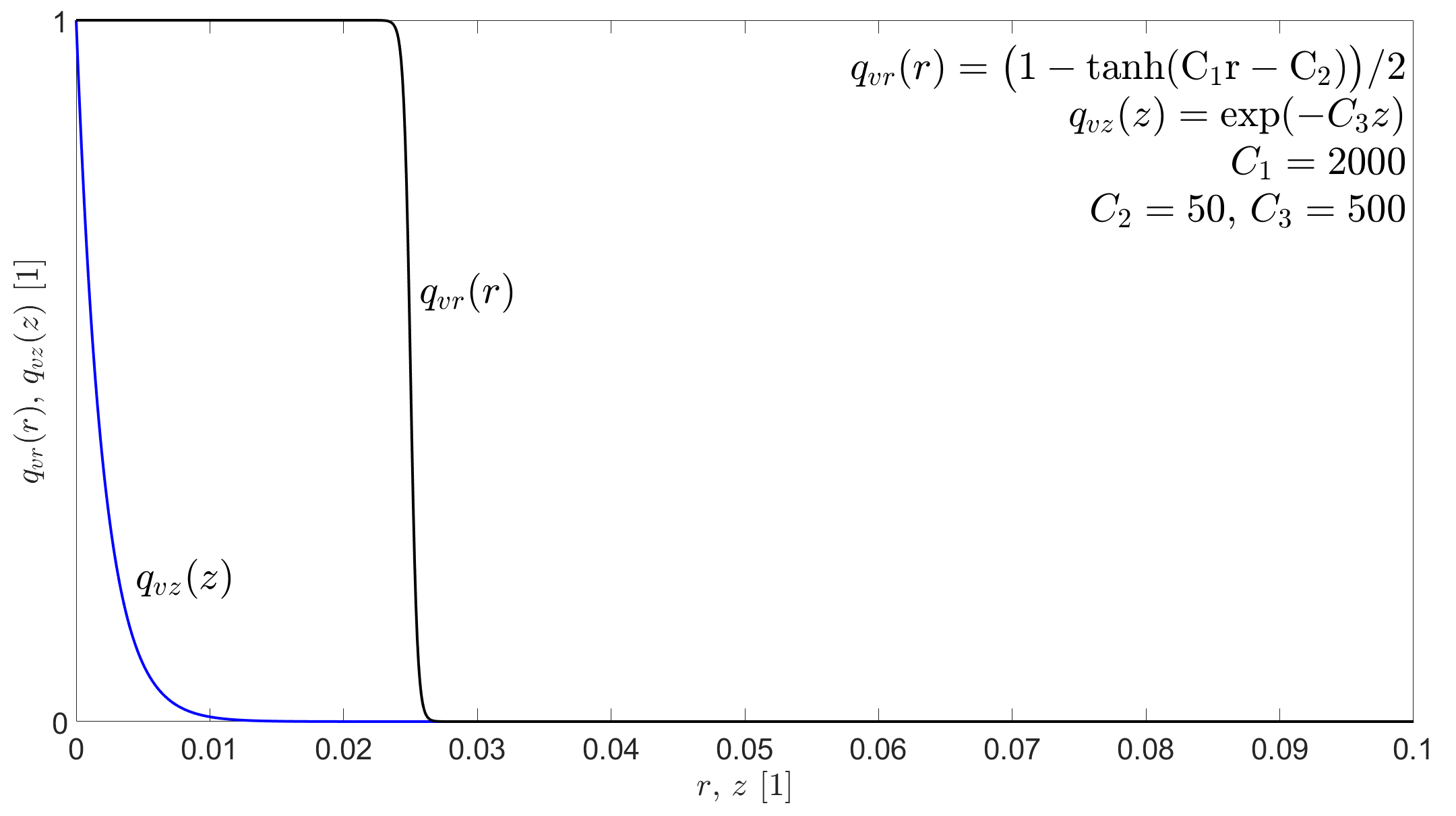}
	\caption{Internal heat generation characteristics (with dimensionless parameters), the heat source is concentrated on the surface with respect to the scaling.}
	\label{fig3}
\end{figure}

Let us consider a plane wall constantly heated on one the left side in a circular area ($r<r_h$) with $q_w$ such as Figure \ref{fig2} (left side) prescribes. In fact, the book of Carslaw and Jaeger \cite{CarJae59b} offers a solution for the Fourier heat equation for constant heating in the domain $r<r_h$, however, we are looking for the temperature history further away from $r_h$ such as how the blue dots showing in Fig.~\ref{fig2}. Additionally, since the boundary condition is space-dependent on the left side, it is challenging and difficult to apply the findings for non-Fourier heat equations. Therefore, we decided to reduce this original problem to a simpler one and substitute the surface heating with a space-dependent, surface-concentrated internal heat generation, as the characteristics present in Fig.~\ref{fig3}. Hence the boundary conditions remain homogeneous but still applicable to the original problem. Furthermore, we assume that at a large enough distance from the heat source, the temperature remains constant, thus we prescribe constant temperature boundary conditions on the right side and on the top.

It is more convenient to formulate our models in a cylindrical coordinate system using dimensionless quantities. For the length scale, we use $R$, the radius of the whole domain, and we set the thickness to be the same, i.e., $L=R$. The usual Fourier number is introduced using the thermal diffusivity $\alpha=\lambda/(\rho c_v)$ for the time scale. The temperature field is homogenized and normalized with the initial temperature $T_0$. Overall, these lead to the following set of dimensionless quantities, 
\begin{align}
	\hat r = \frac{r}{R}, \quad \hat z = \frac{z}{R}, \quad \hat t =  \frac{\alpha t}{R^2}, \quad \hat T = \frac{T - T_0}{T_0}, \quad \hat q_{r,z} = q_{r,z} \frac{R}{\lambda T_0}, \quad  \hat q_v = q_v \frac{R^2}{\lambda T_0}, 
\end{align}	
and thus, the non-Fourier parameters read
\begin{align}	
	\hat \eta_{1,2} = \eta_{1,2} \frac{1}{R^2}, \quad \hat \tau = \frac{\alpha \tau}{R^2}, \quad \hat \tau_T =\frac{\alpha \tau_T}{R^2} .
\end{align}
In the following, we leave the hat notation for simplicity and show the units wherever necessary. 

Taking account that it is a two-dimensional problem for $r$ and $z$ in a cylindrical coordinate system, the balance of internal energy
\begin{align}
	\partial_t T + \partial_r q_r + \frac{1}{r} q_r + \partial_z q_z = q_v(r,z), \quad t\in [0, \infty), \quad (r,z)\in [0,1]\times[0,1],
\end{align}
and the constitutive equations are
\begin{align}
	\textrm{Fourier:} \quad &  q_r = - \partial_r T, \\ 
	\quad & q_z = - \partial_z T, \\
	\textrm{GK:} \quad & \tau \partial_t q_r + q_r = - \partial_r T + (\eta_1 + \eta_2) \left[ \partial_{rr}  - \frac{1}{r^2} + \frac{1}{r} \partial_r \right] q_r + \eta_1 \partial_{zz} q_r + \eta_2 \partial_{rz} q_z, \label{eq7} \\
	\quad & \tau \partial_t q_z + q_z = - \partial_z T + (\eta_1 + \eta_2) \partial_{zz} q_z + \eta_1 \left[\partial_{rr} + \frac{1}{r} \partial_r  \right] q_z  + \eta_2 \left[ \frac{1}{r} \partial_z + \partial_{rz}  \right] q_r, \label{eq8} \\
	\textrm{JE:} \quad & \tau \partial_t q_r + q_r = - \partial_r T - \tau_T \partial_{tr} T, \\
	\quad & \tau \partial_t q_z + q_z = - \partial_z T - \tau_T \partial_{tz} T,
\end{align}
accompanying the $T=0$ initial condition, and $q=0$ and $T=0$ boundary conditions with respect to Fig.~\ref{fig2}. We are looking for the corresponding heat flux and temperature fields.

\section{Solution method}
Although the problem seems complicated, there is quite efficient method to handle such a complicated set of partial differential equations. Here is the strategy we follow. First, we exploit that the problem can be separated into two parts, viz., we can solve the homogeneous ($q_v=0$) transient case ($T_h(r,z,t)$) separately and the inhomogeneous ($q_v\neq 0$) steady-state case ($ T_{\textrm{st}}(r,z)$), and we find the solution as their superposition, $T(r,z,t) = T_{\textrm{st}}(r,z) + T_h(r,z,t)$.
Second, we start with the Fourier heat equation, not merely because we want to compare the non-Fourier solutions to Fourier's, but because we can exploit Fourier's steady-state solution in solving the non-Fourier models. The steady temperature field remains the same for both Fourier and non-Fourier cases. However, the heat flux fields can differ, hence we also include this aspect. 
In the case of Fourier's heat equation, we can quickly determine the proper eigenfunctions and eigenvalues through the Sturm-Liouville problem. It is worth noting that even the non-Fourier models do not introduce higher-order spatial derivatives for the temperature field beyond the Laplacian. Therefore what eigenfunctions we find can also be applied to the GK and JE models.
Third, we solve the non-Fourier models exploiting the ansatz that their solutions can be represented in the same set of eigenfunctions with time-dependent coefficients, this is a sort of Galerkin method. That approach simplifies the complicated system of partial differential equations to a set of ordinary differential equations for the time-dependent coefficients. 

\subsection{Fourier heat equation}
In the case of the Fourier equation, it is easier to start with its $T$-representation, i.e.,
\begin{align}
	\partial_t T = \partial_r^2T + \frac{1}{r} \partial_r T + \partial_z^2 T + q_v(r,z), \label{eqft}
\end{align}
and applying the standard separation of variable technique for the homogeneous part (where $q_v$ is absent), $T_h(r,z,t) = \varphi(t) \xi(r,z)$, one obtains 
\begin{align}
	\textrm{for time:} \quad & \frac{\textrm{d} \varphi}{\textrm{d} t} + \beta^2 \varphi = 0, \\
	\textrm{for space:} \quad & \frac{1}{r} \partial_r \xi + \partial_r^2 \xi + \partial_z^2 \xi + \beta^2 \xi =0, \quad \Rightarrow \quad \xi(r,z) = \rho(r) \zeta(z), \quad \Rightarrow \\
	\textrm{for $r$:} \quad & \frac{1}{r} \frac{\textrm{d} \rho}{\textrm{d} r} +  \frac{\textrm{d}^2 \rho}{\textrm{d} r^2 } + \mu^2 \rho = 0, \\
	\textrm{for $z$:} \quad & \frac{\textrm{d}^2 \zeta}{\textrm{d} z^2 } + \gamma^2 \zeta = 0,
\end{align}
thus $\mu^2 + \gamma^2 = \beta^2$. Applying the boundary conditions 
\begin{align}
	q_r (r=0, z, t) = \partial_r T|_{r=0} = 0, \quad T(r=1,z, t)=0, \quad q_z(r,z=0,t)=\partial_z T|_{z=0}=0, \quad T(r,z=1,t)=0,
\end{align}
we find the following eigenfunctions and eigenvalues,
\begin{align}
	\rho(r) = J_0(\mu_n r), \quad \mu_n: \ J_0(\mu_n)=0, \quad \zeta(z)=\cos(\gamma_m z), \quad \gamma_m = \frac{\pi}{2} + m\pi,
	\label{eig}
\end{align}
and
hence the solution is 
\begin{align}
	T_h(r,z,t) = \sum_{n=1}^\infty \sum_{m=0}^\infty K_{nm} e^{-\beta_{nm}^2 t} J_0(\mu_n r) \cos(\gamma_m z), \quad \beta_{nm}^2 = \mu_n^2 + \gamma_m^2.
\end{align}
Consequently, we can use Eq.~\eqref{eig} to construct the steady-state solution $T_\textrm{st}$, including the heat generation as well, i.e.,
\begin{align}
	T_\textrm{st}(r,z) = \sum_{n=1}^\infty \sum_{m=0}^\infty C_{nm} J_0(\mu_n r) \cos(\gamma_m z), \quad q_v(r,z) = \sum_{n=1}^\infty \sum_{m=0}^\infty B_{nm} J_0(\mu_n r) \cos(\gamma_m z), \label{eq5}
\end{align}
\begin{align}
	B_{nm} = \int\displaylimits_0^1 \int\displaylimits_0^1 r q_{vr}(r) q_{vz}(z) J_0(\mu_n r) \cos(\gamma_m z) \textrm{d}r \textrm{d}z.
\end{align}
Substituting Eq.~\eqref{eq5} into Eq.~\eqref{eqft} (with $\partial_t T = 0$), we can find the relation between the known $B_{nm}$ and the unknown $C_{nm}$, 
\begin{align}
	C_{nm} \left(-\frac{1}{r}  \mu_n J_1 (\mu_n r) - \frac{\mu_n^2}{2}  \Big(J_0(\mu_n r) - J_2(\mu_n r)\Big) - J_0(\mu_n r) \gamma_m^2 \right ) \cos(\gamma_m z) = B_{nm} J_0(\mu_n r) \cos(\gamma_m z). \label{eq6}
\end{align}
Then Eq.~\eqref{eq6} is multiplied with $r J_0(\mu_n r)  \cos(\gamma_m z)$ and integrated from $0$ to $1$ with respect to both $r$ and $z$, following the Galerkin procedure. The $z$-direction is straightforward as both sides are multiplied with  $\cos(\gamma_m z)$, the non-trivial part originates from the $r$ direction, and results in 
\begin{align}
	C_{nm} = B_{nm} \frac{1}{\beta_{nm}^2}, \label{eq10}
\end{align}
which holds for any internal heat generation $q_v(r,z)$. Due to the separation $T(r,z,t) = T_{\textrm{st}}(r,z) + T_h(r,z,t)$, the initial condition for the homogeneous part reads $T_h(r,z,t=0) = -  T_{\textrm{st}}(r,z)$ as $T(r,z,t=0)=0$, and thus $K_{nm}=-C_{nm}$.

\subsection{Guyer-Krumhansl heat equation}
Although we follow the same technique here, we also separate the homogeneous and inhomogeneous parts, but it is more advantageous to determine the steady-state heat flux field first. Since the steady temperature field $T_\textrm{st}(r,z)$ is inherited, we can substitute $\partial_r T_\textrm{st}(r,z)$ and $\partial_z T_\textrm{st}(r,z)$ into Eqs.~\eqref{eq7}-\eqref{eq8}. Furthermore, we suppose that each term inherits the corresponding set of eigenfunctions and eigenvalues as the boundary conditions remain, and thus
\begin{align}
	q_r = \sum_{n=1}^\infty \sum_{m=0}^\infty D_{nm} J_1(\mu_n r) \cos(\gamma_m z), \quad q_z= \sum_{n=1}^\infty \sum_{m=1}^\infty  E_{nm} J_0(\mu_n r) \sin(\gamma_m z), \label{eq9}
\end{align}
respecting the corresponding derivatives, too. In that steady-state, we consider that $\partial_t q_r=\partial_t q_z=0$ in Eqs.~\eqref{eq7}-\eqref{eq8}, and after substituting Eq.~\eqref{eq9} into Eqs.~\eqref{eq7}-\eqref{eq8} and integrating, we obtain a set of algebraic relations among the coefficients,
\begin{align}
	c_1 D_{nm} = \mu_n C_{nm} - c_2 E_{nm}, \quad c_3 E_{nm} = \gamma_m C_{nm} - c_4 D_{nm}, \label{eq11}
\end{align}
with 
\begin{align}
	c_1=1 + (\eta_q+ \eta_2) ( 2+ \mu_n^2) + \eta_1 \gamma_m^2, \quad c_2=\eta_2 \mu_n \gamma_m, \quad c_3=1+\gamma_m^2(\eta_1 + \eta_2) + \mu_n^2 \eta_1, \nonumber \end{align}
\begin{align}
	c_4= 2\gamma_m \eta_2 \left( \frac{1}{\mu_n J_1(\mu_n)^2} +\mu_n  \right). \label{eq12}
\end{align}
Since $C_{nm}$ is known from Eq.~\eqref{eq10}, Eq.~\eqref{eq11} can be solved for $D_{nm}$ and $E_{nm}$, and it holds for any heat sources. Then the steady-state solution is given by Eqs.~\eqref{eq5} and \eqref{eq9}. 

The transient (homogeneous) solution is constructed similarly, however, the coefficients are now time-dependent, viz., we assume that
\begin{align}
	T_h(r,z,t) = \sum_{n=1}^\infty \sum_{m=0}^\infty \tilde C_{nm}(t) J_0(\mu_n r) \cos(\gamma_m z), \quad q_r(r,z,t)=\sum_{n=1}^\infty \sum_{m=0}^\infty \tilde D_{nm}(t) J_1(\mu_n r) \cos(\gamma_m z), \label{eq15}
\end{align}
\begin{align}
	q_z(r,z,t)=\sum_{n=1}^\infty \sum_{m=1}^\infty \tilde E_{nm}(t) J_0(\mu_n r) \sin(\gamma_m z), \label{eq16}
\end{align}
are still valid. Furthermore, now we need to exploit the energy balance as well to obtain $\tilde C_{nm}(t)$, 
\begin{align}
	\frac{\textrm{d}}{\textrm{d}t} \tilde C_{nm}(t) + \mu_n \tilde D_{nm}(t) + \gamma_m \tilde E_{nm}(t) = 0.
\end{align}
After following the same procedure, we obtain almost the same set of equations except that the time derivative terms appear. Consequently, the set of PDEs is reduced to a set of ODE,

\begin{gather}
	\frac{\textrm{d}}{\textrm{d}t} \begin{bmatrix} \tilde C_{nm}(t) \\ \tilde D_{nm}(t) \\ \tilde E_{nm}(t) \end{bmatrix}
	=
	\begin{bmatrix}
		0 & -\mu_n & -\gamma_m \\
		\frac{\mu_n}{\tau} & -\frac{c_1}{\tau} & -\frac{c_2}{\tau} \\
		\frac{\gamma_m}{\tau} & -\frac{c_4}{\tau} & - \frac{c_3}{\tau}
	\end{bmatrix}
	\begin{bmatrix} \tilde C_{nm}(t) \\ \tilde D_{nm}(t) \\ \tilde E_{nm}(t) \end{bmatrix}, \label{eq13}
\end{gather}
in which the coefficients are inherited from Eq.~\eqref{eq12}, and its solution can easily be found in the form of a matrix exponential, $\exp(\mathbf M_{nm} t)$. Let us recall that the initial condition $T_h(r,z,t=0) = -  T_{\textrm{st}}(r,z)$ is the same, and therefore we can exploit the known coefficients $\tilde C_{nm}(t=0)=-C_{nm}$, $\tilde D_{nm}(t=0)=-D_{nm}$, and $\tilde E_{nm}(t=0)=-E_{nm}$.

\subsection{Jeffreys heat equation}
Here, the solution is simpler as the steady-state heat flux field is identical to Fourier's case, hence we do not need to compute it separately. We repeat the ansatz of Eq.~\eqref{eq15}-\eqref{eq16} to determine the coefficient matrix $\mathbf M_{nm}$, and following the same steps, that procedure results in
\begin{gather}
	\frac{\textrm{d}}{\textrm{d}t} \begin{bmatrix} \tilde C_{nm}(t) \\ \tilde D_{nm}(t) \\ \tilde E_{nm}(t) \end{bmatrix}
	=
	\begin{bmatrix}
		0 & -\mu_n & -\gamma_m \\
		\frac{\mu_n}{\tau} & -\frac{1+\tau_T\mu_n^2}{\tau} & -\frac{\tau_T \mu_n \gamma_m}{\tau} \\
		\frac{\gamma_m}{\tau} & -\frac{\tau_T \mu_n \gamma_m}{\tau} & -\frac{1+\tau_T\gamma_m^2}{\tau}
	\end{bmatrix}
	\begin{bmatrix} \tilde C_{nm}(t) \\ \tilde D_{nm}(t) \\ \tilde E_{nm}(t) \end{bmatrix}, \label{eq14}
\end{gather}
together with the known coefficients from Fourier's solution, the time evolution for the Jeffreys case is obtained in the form of $\exp(\mathbf M_{nm} t)$.

\section{Steady-state distributions}
First, let us begin with the steady-states, primarily focusing on the differences between the Fourier and GK equations. From a theoretical point of view, and according to Alvarez et al.~\cite{AlvEtal12}, GK's steady heat flux field can differ from Fourier's. Here we have to discover in what sense and in what measure they can differ from each other. From a practical point of view, it is possible to measure the average local heat flux, usually on a minimal area of $10\times10$ mm$^2$ up to about $80\times80$ mm$^2$. Furthermore, it is known that such sensors can significantly distort the local heat flux field \cite{KissL93, KissBui99}. Consequently, if one aims to observe the traces of non-Fourier heat conduction at this level, such spatial scales must be included in the preliminary analysis as smooth flux distributions (such as for the temperature field) cannot be measured. These analytical calculations can ease such analysis as well. Since the JE model has the same steady-state as Fourier's, we leave this analysis aside for that heat equation. 

\begin{figure}[]
	\centering
	\includegraphics[width=17cm,height=6cm]{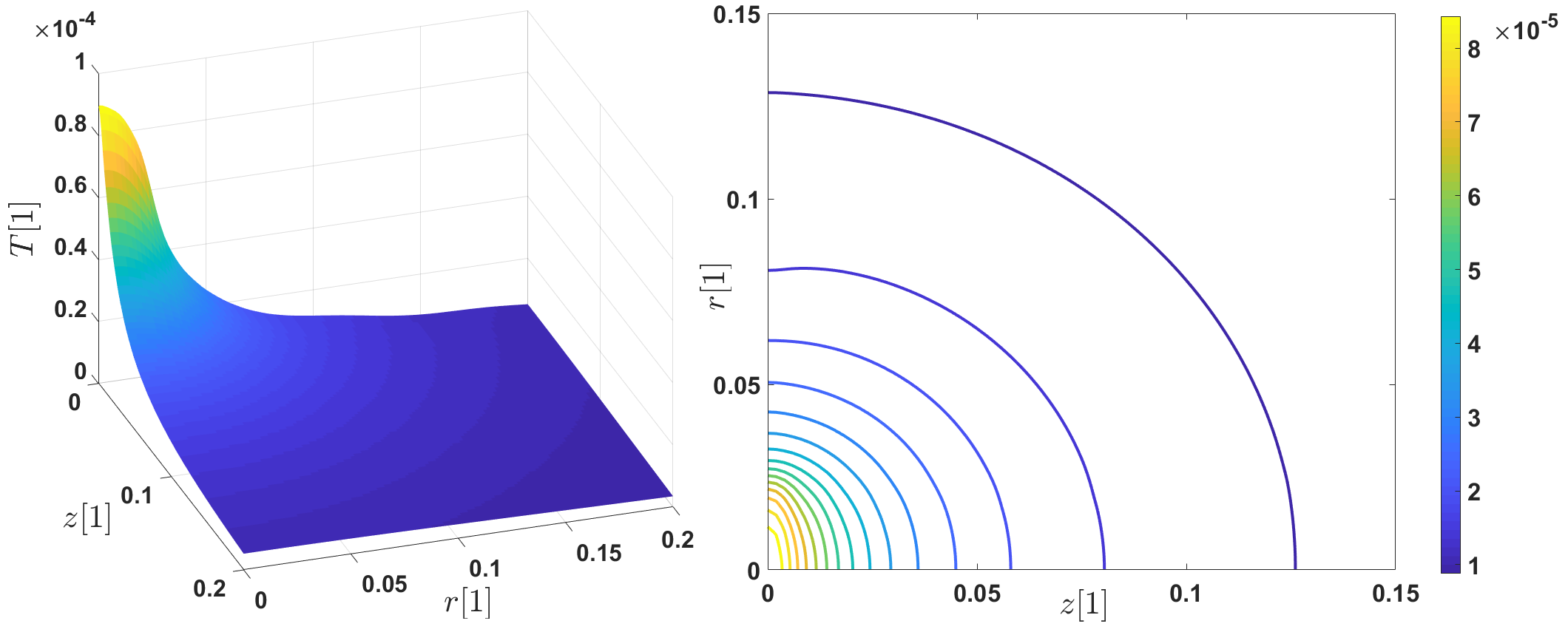}
	\caption{Two-dimensional steady-state temperature distribution, using $N=M=200$ terms, showing only partial spatial domain.}
	\label{fig4}
\end{figure}

\begin{figure}[]
	\centering
	\includegraphics[width=17cm,height=5.5cm]{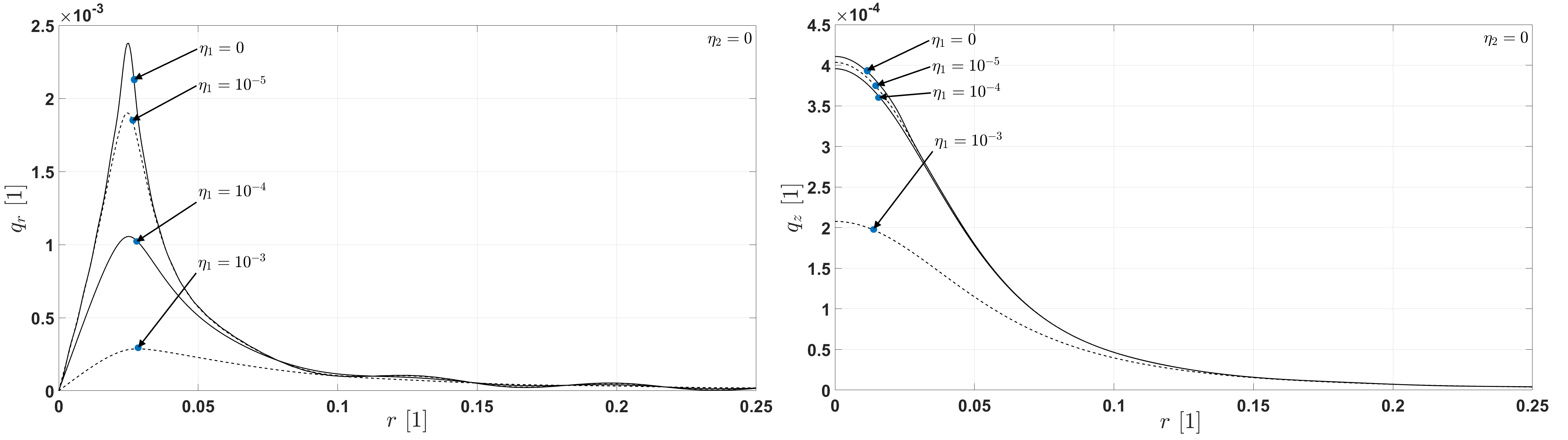}
	\caption{Steady-state $q_r(r,z=0)$ (left) and $q_z(r,z=0.05)$ (right) distributions with $\eta_2=0$, using $N=M=200$ terms, showing only partial spatial domain.}
	\label{fig5}
\end{figure}

\begin{figure}[]
	\centering
	\includegraphics[width=17cm,height=5.5cm]{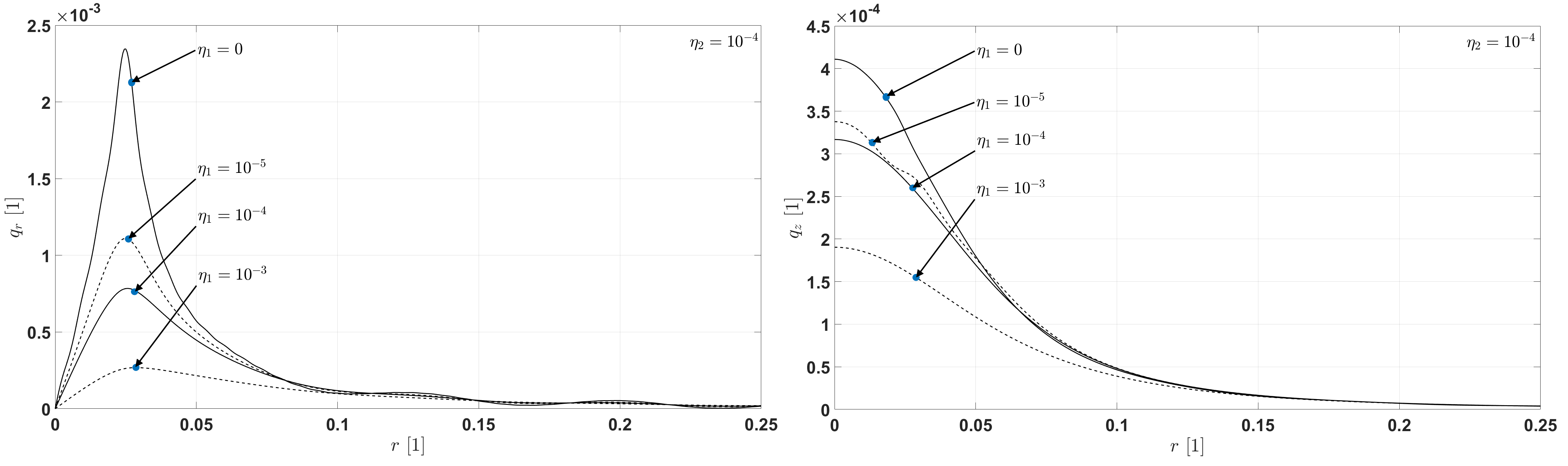}
	\caption{Steady-state $q_r(r,z=0)$ (left) and $q_z(r,z=0.05)$ (right) distributions with $\eta_2=10^{-4}$, using $N=M=200$ terms, showing only partial spatial domain.}
	\label{fig6}
\end{figure}

Fig.~\ref{fig4} presents the temperature distribution for Fourier's heat equation, which remains the same for the Guyer-Krumhansl and Jeffreys equations. Concerning the heat flux field, the situation becomes quite different. It is worth studying the outcome of the GK equation closer, see Fig,~\ref{fig5} for the details about $q_r(r,z=0)$ and $q_z(r,z=0.05)$. That characteristics is preserved for any $q_r(r,z=\textrm{const.})$ distributions. The influence of $\eta_1$ is clear, and it can significantly decrease the maximum. Similarly to $\eta_1$, $\eta_2$ has the same effect on the heat flux field, being more influential on $q_z$, see Fig.~\ref{fig6} for a particular solution. 

Although these effects are clear and strong for such a parameter interval, the situation of observing them in a steady-state is not that hopeful. Let us consider the flash experiments on rocks and their GK-evaluation with a one-dimensional model \cite{FehKov21}, we find that $\eta_1 + \eta_2 \approx 10^{-7}$ m$^2$, in general. Consequently, the most substantial effects could be observed for a body with $R=0.01$ m or less, which would probably violate our initial assumptions, and the boundary conditions would not be valid as well. For larger bodies, e.g., with $R=0.1$ m or even larger ($R=1$ m), the effect becomes small and difficult to detect. This is one reason why this scaling property is most important for nanoscale objects.

\section{Transient distributions}
\subsection{Fourier's heat equation.} 
\begin{figure}[]
	\centering
	\includegraphics[width=14cm,height=13cm]{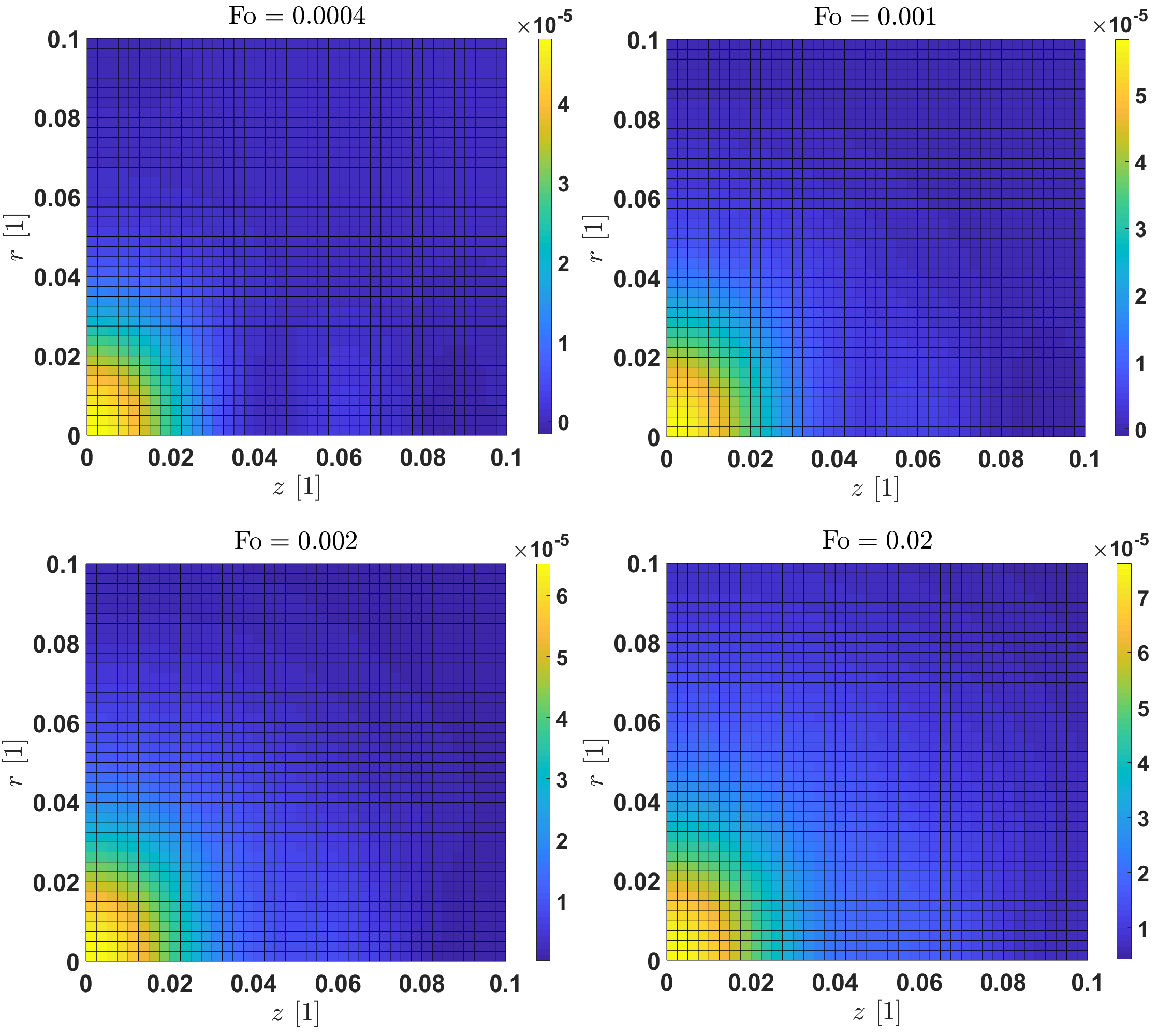}
	\caption{Transient temperature distribution following from Fourier's heat equation, using $N=M=40$ terms, showing only partial spatial domain.}
	\label{fig7}
\end{figure}

\begin{figure}[]
	\centering
	\includegraphics[width=7cm,height=6cm]{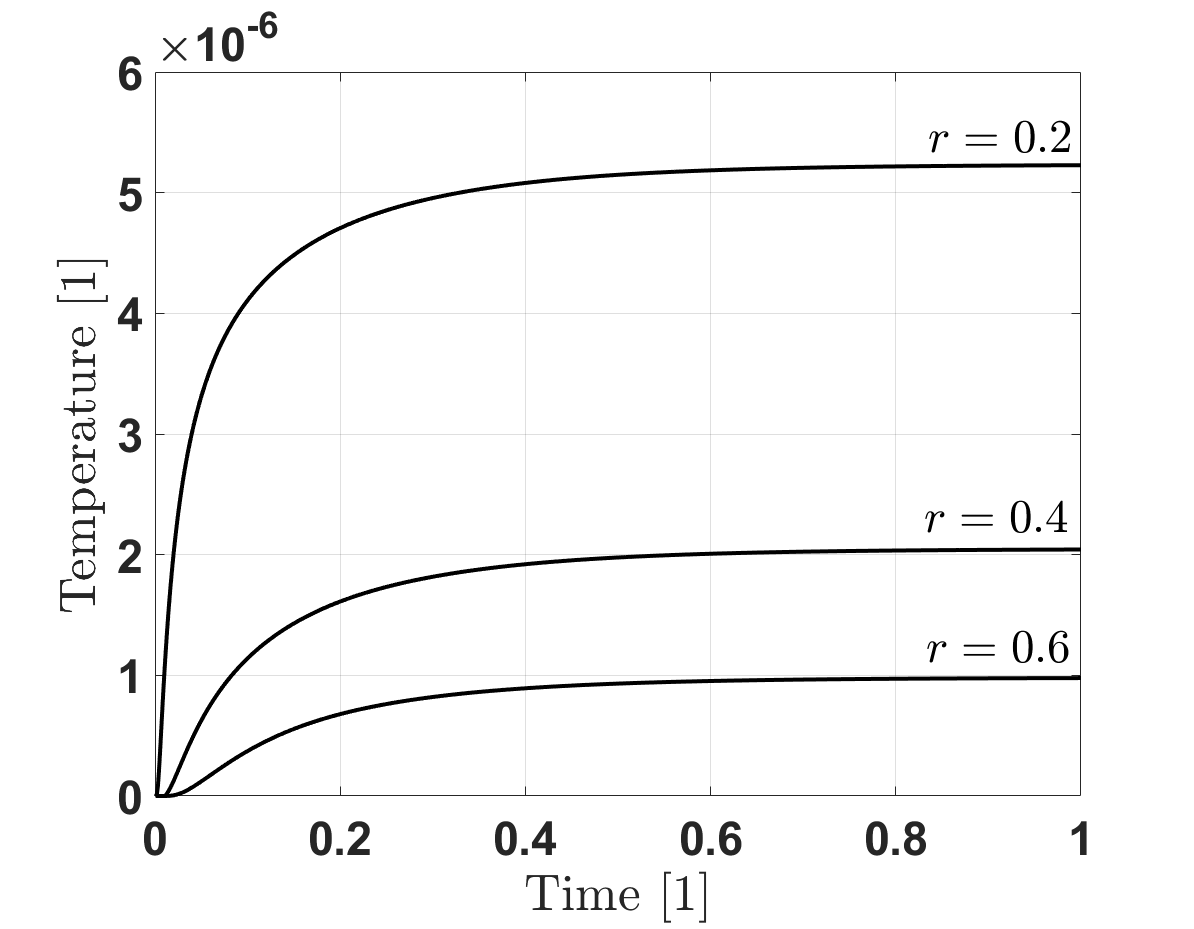}
	\caption{Transient temperature history on the surface at different radii from Fourier's equation, using $N=M=100$ terms.}
	\label{fig8}
\end{figure}

Let us recall that we are using the conventional Fourier number for the time scale. The characteristic size could be $1$ m as relatively large bodies are considered. Consequently, we must choose small Fourier numbers since those can express relatively large time instants. Figure \ref{fig7} shows the temperature distribution for Fourier's heat equation, in which we can observe that the characteristics of the distribution establish quickly and does not change significantly for larger time intervals. The color scaling, however, changes, showing how the equilibrium is approximated. At farther away from the heat source, the temperature changes more slowly since the gradients are much smaller. This is also presented in Fig.~\ref{fig8}, showing the surface temperature history at different radii, as it is denoted previously in Fig.~\ref{fig2}. We use Fig.~\ref{fig8} for comparative purposes in case of non-Fourier models as temperature maps (such as Fig.~\ref{fig7}) would not highlight the characteristics of the non-Fourier behavior.

\subsection{Guyer-Krumhansl heat equation.}
\begin{figure}[]
	\centering
	\includegraphics[width=14cm,height=6cm]{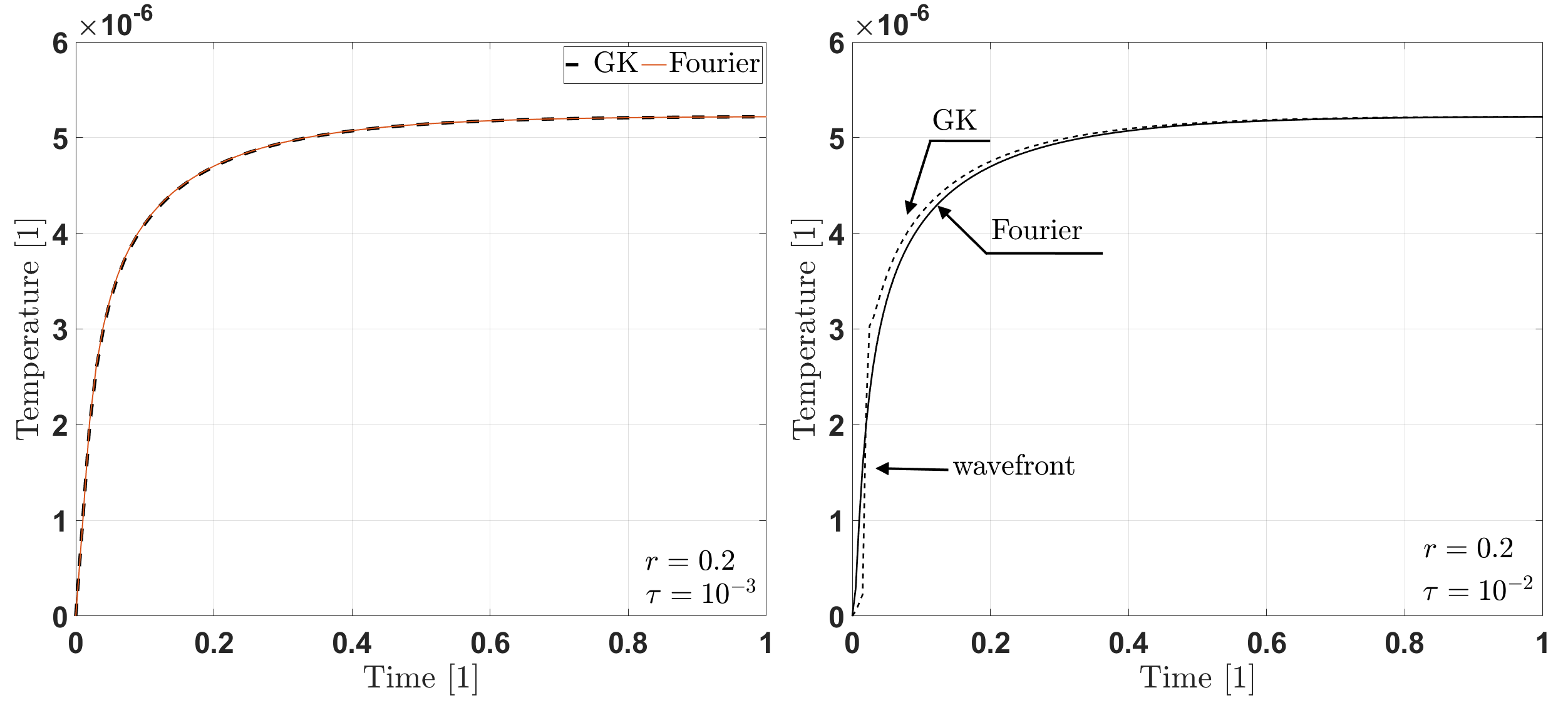}
	\caption{Transient temperature history on the surface, comparing the Fourier and GK equations with  $\eta_1=\eta_2=0$ and $\tau=10^{-3}$ (left), $\tau=10^{-2}$ (right), using $N=M=100$ terms.}
	\label{fig9}
\end{figure}

Although the GK equation can reproduce the heat wave solutions following the Cattaneo equation with $\eta_1=\eta_2=0$, called second sound, it is not meaningful in our situation. First, in order to generate such a heat wave, the time scale of the excitation (e.g., a heat pulse) should match the material's characteristic properties, most importantly, its relaxation time $\tau$. Since its values range from $10^{-10} -- 1$ s depending on the propagation phenomenon we investigate \cite{Strau11b, RuggSug21b, JouEtal96b}, such effects become irrelevant for large bodies on much larger time scales. Second, we use the GK equation as an effective approach to model the parallel diffusive mechanisms instead of modeling heat waves (or anything else related to phonon hydrodynamics). Fig.~\ref{fig9} presents two cases for the $\eta_1=\eta_2=0$ setting, with $\tau=10^{-3}$ (left) and $\tau=10^{-2}$ (right). Similarly to the Fourier number, the relaxation time has the same scaling, therefore even $\tau=10^{-3}$ is so large, it still does not show any difference from Fourier's solution, unlike the second case with unrealistically large relaxation time, but the effect is still weak. This setting will not be relevant for such continuous heating in a large macroscale body. 

\begin{figure}[]
	\centering
	\includegraphics[width=7cm,height=6cm]{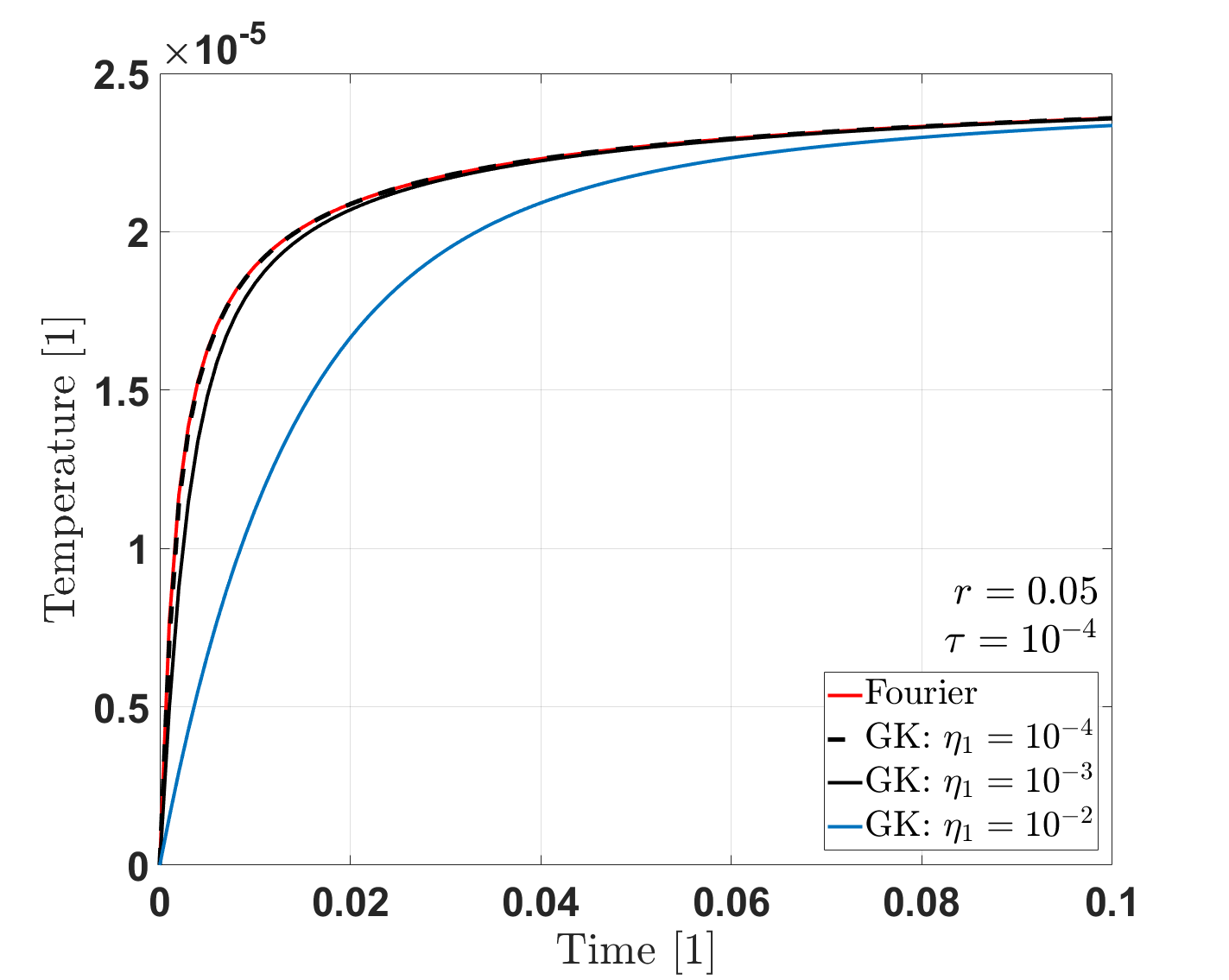}
	\caption{Transient temperature history on the surface, comparing the Fourier and GK equations with $\eta_2=0$, using $N=M=100$ terms.}
	\label{fig10}
\end{figure}

\begin{figure}[]
	\centering
	\includegraphics[width=17cm,height=5.8cm]{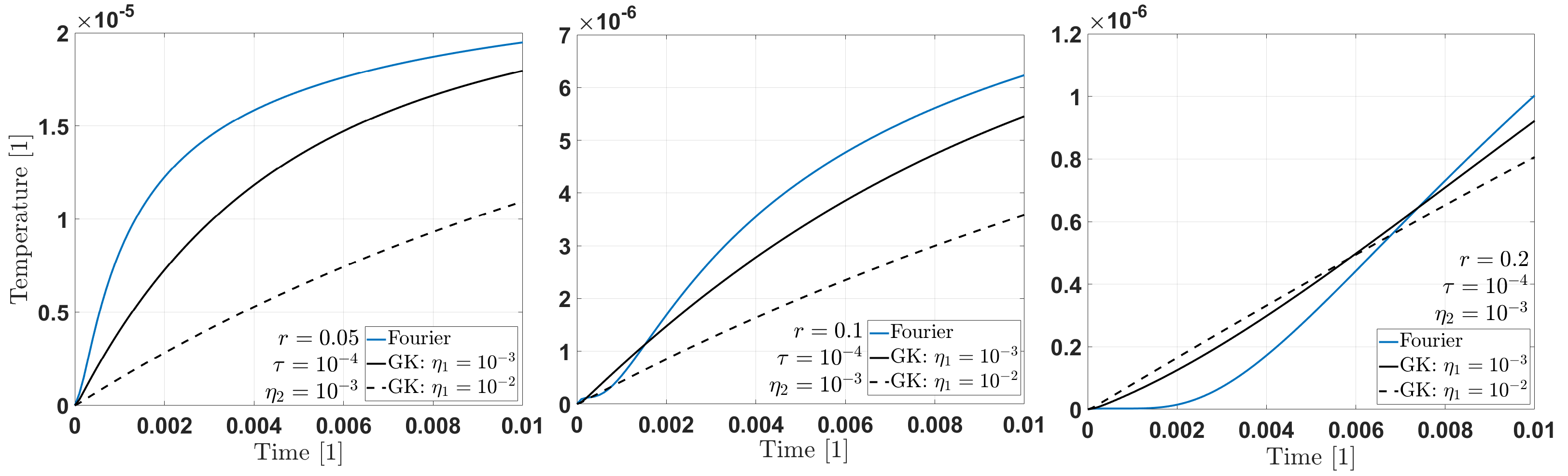}
	\caption{Transient temperature history on the surface at different radii, comparing the Fourier and GK equations, using $N=M=100$ terms.}
	\label{fig11}
\end{figure}

In fact, it is not easy to find parameters resulting in a remarkably different solution. Figure \ref{fig10} presents an example in which the radius we use is much closer to the heat source, for farther away and longer time intervals, the differences vanish. 
It is worth studying how we can recover a similar behavior observed in heat pulse experiments (see Fig.~\ref{fig1} for the experimental characteristics). Interestingly, being closer to the heat source, GK's solution is not faster than Fourier's, however, it changes with the radius. Figure \ref{fig11} presents these characteristics, and thus it is best to measure the temperature farther from the heat source to more reliably observe the non-Fourier behavior.

\subsection{Jeffreys heat equation.}
The $\tau_T=0$ subcase coincides with the previous analysis using the GK equation (see the right side of Fig.~\ref{fig9}), therefore we do not repeat it again. Instead, we focus on studying the effects induced by the extra time derivative term in the constitutive relation with $\tau_T\neq 0$ and comparing our findings with Fig.~\ref{fig11}. Although $\tau_T$ acts analogously on the temperature field, their heat flux fields differ. That difference is essential for superfluids and further low-temperature modeling problems \cite{SellSciAmen19}. In engineering practice, however, that difference could be negligible as we seek only an effective model to provide a more accurate description of heterogeneous materials. An effective description cannot simultaneously model the temperature and heat flux fields. In that sense, neither approach is more accurate than the other. Both can reproduce the same temperature history, however, the GK equation can be much more complicated. On the one hand, for the JE model, as simply the time derivative of Fourier's equation is added, it is easier to utilize the usual approaches for initial and boundary conditions. On the other hand, although thermodynamically compatible, the JE model does not fit into the systematic generalization of non-Fourier equations. 

\begin{figure}[]
	\centering
	\includegraphics[width=17cm,height=5.8cm]{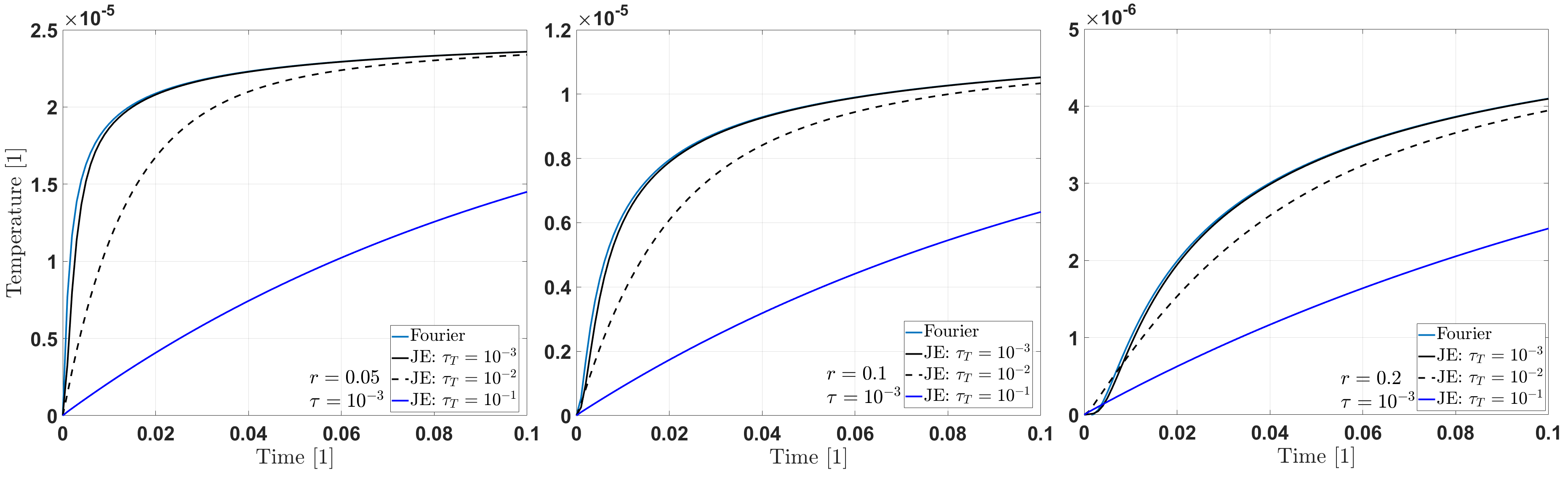}
	\caption{Transient temperature history on the surface at different radii, comparing the Fourier and JE equations, using $N=M=100$ terms.}
	\label{fig12}
\end{figure}

\section{Discussion}
There is a need for an advanced heat conduction model to overcome the difficulties emerging together with the use of complex heterogeneous materials. Their effective thermal behavior can significantly differ from Fourier's prediction, even though the classical heat equation governs each component. The interaction of parallel heat transfer channels results in a non-Fourier behavior. Two promising extensions of the Fourier equation are the Guyer-Krumhansl and the Jeffreys models. Both describe two heat conduction time scales, even in an isotropic setting. These models are analytically solved for a two-dimensional situation. 

The analytical solution revealed that there is no need for additional boundary conditions for these GK and JE models, the same set of eigenfunctions can be used. That property notably eases the solution of these more complex models. The resulting ordinary differential equations can be solved easily for the linear situation; thus, unknown time-dependent coefficients are found. Furthermore, we also exploited that the steady-state temperature distribution given by Fourier's law remains the same in the non-Fourier case, therefore, it is advantageous to handle that heat conduction problem as a superposition of the transient and steady distributions. It also makes it more straightforward how to take into account the initial conditions. 

Studying the time evolution of the surface temperature time histories $T(t)$, we observed similarities compared to the one-dimensional heat pulse experiments, however, these situations are not directly comparable. First, as the surface temperature is the most straightforward to measure, it seems more advantageous if the thermometer is situated further from the heated region to observe the over-diffusive behavior possibly. This can change significantly in space. Second, with such constant heating, no steep transients occur in a large body due to its large heat capacity, thus, it is more challenging to observe the non-Fourier behavior. Based on the analytical solutions, it is clear that both models require unusually large parameters to obtain any observable deviation from Fourier's law. In the future, it would be worth investigating the outcome of periodic heating, especially the effects of the period time, and that could enhance the over-diffusive phenomenon. Third, for the GK equation, while the steady heat flux field differs from Fourier's, it is not practically measurable due to the steep change in the heated region, the available heat flux sensors are too large for such an application. Moreover, as the most significant difference occurs under the heater, it excludes this possibility. 
However, the analytical calculations also revealed that if such constant heating does not induce notable non-Fourier effects, one could design a measurement method to characterize such objects with Fourier's law remaining valid thermally.

\section*{Funding}

Project no.~TKP-6-6/PALY-2021 has been implemented with the support provided by the Ministry of Culture and Innovation of Hungary from the National Research, Development and Innovation Fund, financed under the TKP2021-NVA funding scheme. The research was funded by the Sustainable Development and Technologies National Programme of the Hungarian Academy of Sciences (FFT NP FTA), by the grant National Research, Development and Innovation Office-NKFIH FK 134277, and supported by the ÚNKP-22-5-BME-312 New National Excellence Program of the Ministry for Culture and Innovation from the source of the National Research, Development and Innovation Fund.

\section*{Declarations}
\textbf{Conflict of interest} The author declares no competing interests.

\vspace{1cm}

\bibliographystyle{unsrt}

\end{document}